\documentclass[aps,prl,twocolumn,superscriptaddress,longbibliography]{revtex4-2}
\pdfoutput=1
\usepackage{graphicx}
\usepackage{amssymb,amsmath}
\usepackage{bm}
\usepackage{dcolumn}
\usepackage{float}
\usepackage[OT1]{fontenc} 
\usepackage{url}
\usepackage{mathrsfs}
\usepackage{slashed,comment}
\usepackage{color}
\usepackage{verbatim}
\usepackage{enumitem}
\usepackage{soul,physics}
\usepackage[driverfallback=dvipdfm]{hyperref}
\hypersetup{pdfpagemode=FullScreen,colorlinks=true,breaklinks,urlcolor=blue,linkcolor=blue,citecolor=blue}

\usepackage{subfigure}
\usepackage{amssymb}
\usepackage{bm}
\usepackage{graphicx}
\usepackage{amsmath,amssymb}

\usepackage{pdfpages} 
\usepackage{pgffor} 

\makeatletter
\AtBeginDocument{\let\LS@rot\@undefined}
\makeatother

\def\supplementfilename{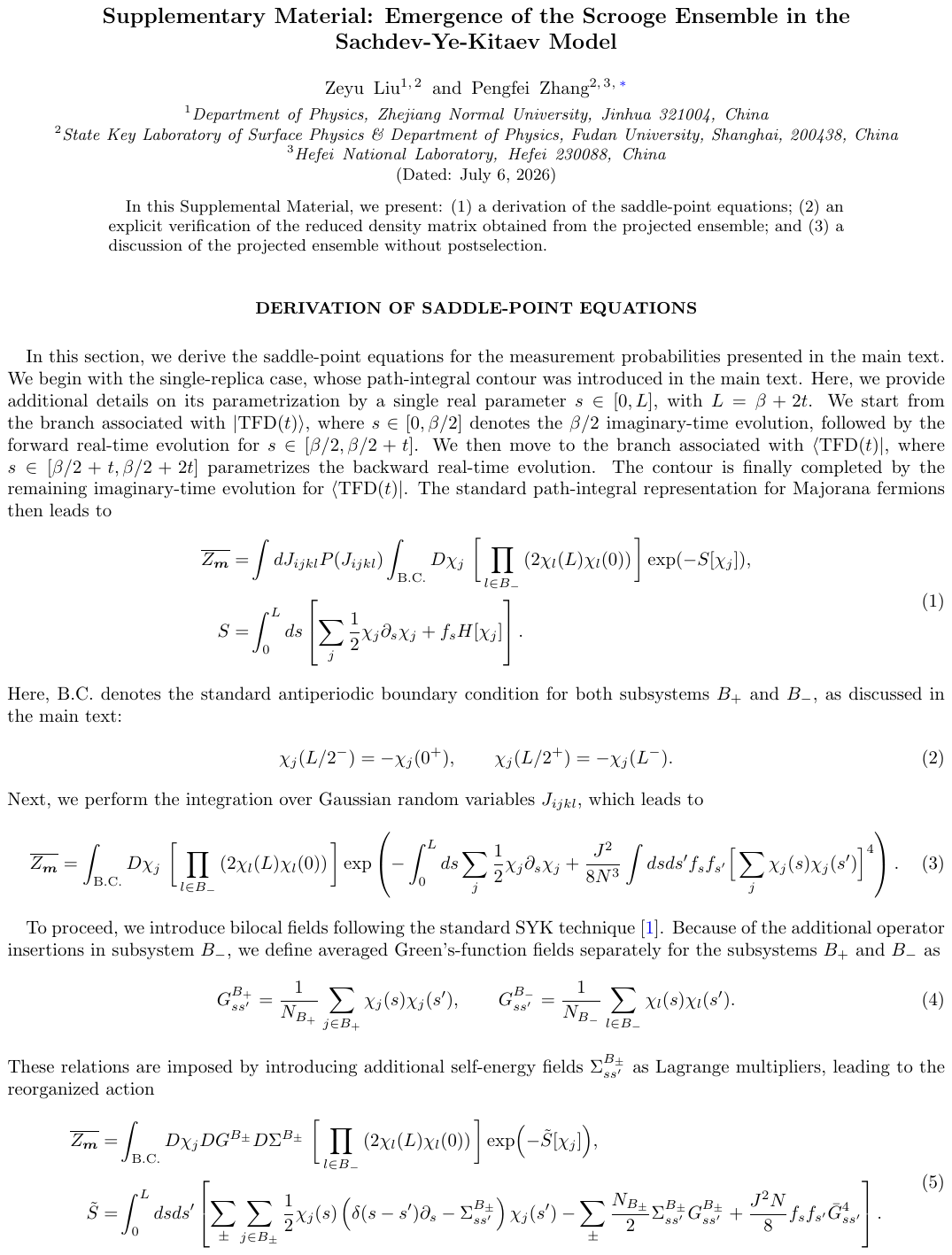}

\pdfximage{\supplementfilename}
\def\numbersupplementpages{\the\pdflastximagepages}

\newif\ifarXiv
\arXivtrue 

\begin{document}
 
  \title{Emergence of the Scrooge Ensemble in the Sachdev-Ye-Kitaev Model }

  \author{Zeyu Liu}
  \affiliation{Department of Physics, Zhejiang Normal University, Jinhua 321004, China}
  \affiliation{State Key Laboratory of Surface Physics \& Department of Physics, Fudan University, Shanghai, 200438, China}

  \author{Pengfei Zhang}
  \thanks{PengfeiZhang.physics@gmail.com}
  \affiliation{State Key Laboratory of Surface Physics \& Department of Physics, Fudan University, Shanghai, 200438, China}
  \affiliation{Hefei National Laboratory, Hefei 230088, China}

  \date{\today}

  \begin{abstract}
  The probabilistic nature of quantum measurement provides a direct window into the structure and complexity of many-body wave functions. When only part of a system is measured, the remaining degrees of freedom form an ensemble of post-measurement states whose statistical structure can reveal a stronger form of thermalization, known as deep thermalization. Recent numerical evidence suggests that this phenomenon is characterized by convergence of the projected ensemble to the Scrooge ensemble, a maximally random ensemble compatible with a given density matrix. In this Letter, we use the solvable Sachdev-Ye-Kitaev (SYK) model to unveil the mechanism by which the Scrooge ensemble emerges in many-body systems. By formulating measurement probabilities and post-measurement states in terms of path integrals, we analytically characterize all moments of the projected ensemble and show that they exactly match those of the Scrooge ensemble, even at short evolution times. We further connect this result to the saddle-point structure of the measurement path integral, which naturally generates the replica permutations underlying Scrooge statistics. Our results establish the solvable SYK model as a tractable setting for exploring universal statistics of quantum measurements in chaotic many-body dynamics.
  \end{abstract}
  \maketitle

  \emph{ \color{blue}Introduction.---} Quantum measurement plays a central role in modern quantum science and technology, where increasingly controllable many-body systems are probed through large sets of measurement outcomes. In a fixed measurement basis, each outcome samples the wave function at the level of an individual configuration. The resulting probability distribution therefore contains fine-grained information about the buildup of complexity and the emergence of statistical behavior under chaotic dynamics~\cite{deutsch1991quantum,srednicki1996thermal,Susskind:2014rva,Stanford:2014jda}. When many degrees of freedom are measured, this distribution can be used to diagnose anticoncentration, namely the delocalization of measurement probabilities over an exponentially large set of outcomes~\cite{Dalzell:2020vxs,Boixo:2016vge,Fefferman:2024dbf,RevModPhys.95.035001,Turkeshi:2024wgi,PhysRevLett.134.050405,PhysRevLett.134.010401,Magni:2025zqf,Christopoulos:2024hii,Magni:2025tch,Tirrito:2024kts,Sauliere:2025zww,Sauliere:2025bpr}. A natural diagnostic is the participation entropy~\cite{PhysRevLett.112.057203,PhysRevLett.123.180601,PhysRevLett.128.130605}, which measures the typical spread of measurement probabilities and is closely related to the classical hardness of sampling problems in quantum information theory~\cite{aaronson2011computational,PhysRevLett.117.080501,Bouland:2018bva,PRXQuantum.3.020328}.

  An even richer situation arises when only part of the system is measured. Such measurements do not merely produce probabilities: they also generate ensembles of post-measurement states on the unmeasured subsystem~\cite{Choi:2021npc}. This observation motivates the notion of deep thermalization, which asks whether the resulting projected ensembles exhibit ergodicity at the level of higher moments~\cite{Claeys2022Biunitarity,Ippoliti2022SolvableDeepThermalization,Ho2022ExactEmergent,Choi2023PreparingRandomStates,Cotler2023EmergentDesigns,Ippoliti2023DynamicalPurification,Liu2024GaussianDeepThermalization,Mark2024MaximumEntropy,Chang2025ChargeConserving,mcginley2025scroogeensemblemanybodyquantum,Feng2026ResourceLocalizability,Liu2026CoherenceInduced,Mok2026NatureStingy}. For systems with a nontrivial average density matrix, the relevant random-state ensemble is the Scrooge ensemble~\cite{PhysRevA.49.668,Goldstein:2005ntv,2008JSP...132..921R,Goldstein:2015syr}, which generalizes the Haar ensemble by incorporating the coarse-grained constraint set by the average state while remaining maximally random. Nevertheless, the underlying mechanism by which the Scrooge ensemble emerges in Hamiltonian systems remains less understood, largely due to the absence of concrete solvable examples.

  \begin{figure}[t]
    \centering
    \includegraphics[width=0.7\linewidth]{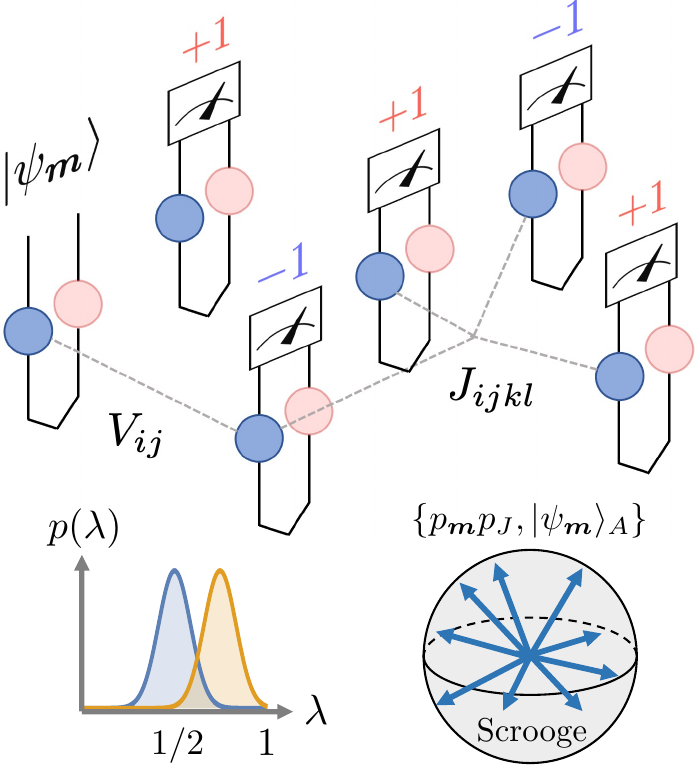}
    \caption{Schematic illustration of the main setup. The system is initialized in a thermofield double state of $N$ pairs of Majorana fermions, denoted by the blue and red circles. After time evolution, projective measurements are performed on most pairs of Majorana fermions, yielding outcomes $\bm{m}$ and a post-measurement state $\ket{\psi_{\bm{m}}}_A$. We analyze the distribution $p(\lambda)$ of measurement outcomes with $\lambda N$ entries of $+1$, as well as the projected ensemble $\{p_{\bm{m}}p_{J},\ket{\psi_{\bm{m}}}_A\}$, where $p_{J}$ is the distribution of random couplings. }
    \label{fig:schematics}
  \end{figure}

  In this Letter, we unveil the mechanism by which the Scrooge ensemble emerges by investigating quantum measurements in Sachdev-Ye-Kitaev (SYK) models~\cite{Sachdev:1992fk,kitaev2014talk,maldacena2016remarks,RevModPhys.94.035004}. As sketched in Fig.~\ref{fig:schematics}, the setup consists of preparing the system in a thermofield double state~\cite{Israel:1976ur,Maldacena:2001kr}, evolving it unitarily, and then performing projective measurements on most Majorana fermions. By formulating measurement probabilities in terms of path integrals, we obtain the full measurement-outcome distribution and thereby enable a controlled analysis of participation-entropy dynamics. More importantly, this framework allows us to analytically determine all moments of the projected ensemble. We show that these moments exactly coincide with those of the Scrooge ensemble, even at arbitrarily short times, revealing the saddle-point mechanism underlying its emergence. These results establish solvable SYK models as a tractable platform for analyzing quantum measurements and their emergent statistical structures under chaotic dynamics.

  \emph{ \color{blue}Setup.---} We consider a many-body system of $N$ Majorana fermions $\chi_j$, with $j\in\{1,2,\cdots,N\}$. The Majorana operators satisfy the canonical anticommutation relations $\{\chi_j,\chi_k\}=\delta_{jk}$. The system is governed by the SYK Hamiltonian:
  \begin{equation}\label{eq:H}
  H=\sum_{j<k<l<m}J_{jklm}\chi_j\chi_k\chi_l\chi_m.
  \end{equation}
  The random couplings $J_{jklm}$ are independent Gaussian variables with zero means and variances $\overline{J_{jklm}^2}=6J^2/N^3$. This scaling enables a controlled field-theoretic analysis via the large-$N$ expansion~\cite{maldacena2016remarks}, making it a paradigmatic solvable model for studying chaotic quantum dynamics. In particular, measurement effects have been analyzed in Refs.~\cite{Jian:2021hve,Antonini:2022lmg}. Nevertheless, these studies either consider forced measurements or focus on setups in which all measurement outcomes occur with equal probability.

  We initialize the system by entangling the Majorana fermions with $N$ auxiliary modes $\eta_j$ in a thermofield double (TFD) state~\cite{Israel:1976ur,Maldacena:2001kr}, a construction with broad applications in condensed matter physics~\cite{Liu:2024mme,Weinstein:2024fug}, quantum information~\cite{Gao:2016bin,Maldacena:2017axo,Susskind:2017nto,Gao:2018yzk,Brown:2019hmk,Gao:2019nyj,Schuster:2021uvg,Jafferis:2022crx,Zhou:2024osg,Liu:2024nhs}, and high-energy physics~\cite{Maldacena:2001kr,Burrage:2018pyg,Kading:2022jjl}. To define the TFD state, we first introduce the maximally entangled state through the conditions $2i\chi_j\eta_j\lvert\mathrm{EPR}\rangle=\lvert\mathrm{EPR}\rangle$ for all $j$. The TFD state then reads
  \begin{equation}
  \lvert\mathrm{TFD}\rangle=e^{-\frac{\beta H}{2}}\lvert\mathrm{EPR}\rangle/\sqrt{Z_\beta}.
  \end{equation}
  Here, $Z_\beta=\mathrm{Tr}_\chi(e^{-\beta H})$ is the partition function of the original $\chi$ system, which ensures the normalization of the TFD state. We then evolve the system to time $t$ under the Hamiltonian \eqref{eq:H}, yielding $ \lvert\mathrm{TFD}(t)\rangle= e^{-iHt}\lvert\mathrm{TFD}\rangle$. 

  Next, we perform quantum measurements on the system. We now divide the system into two subsystems, $A$ and $B$, where $A$ contains a small number $N_A\sim O(1)$ of fermion pairs and $B$ contains the remaining $N_B$ pairs. We projectively measure subsystem $B$ with respect to the operators $2i\chi_j\eta_j$, obtaining outcomes $m_j=\pm1$ for $j\in B$. The corresponding measurement record is denoted by $\bm{m}$, and the product state on $B$ is $\lvert\bm{m}\rangle_B=\otimes_{j\in B}\lvert m_j\rangle_j$, with $2i\chi_j\eta_j\lvert m_j\rangle_j=m_j\lvert m_j\rangle_j$. The maximally entangled state used to construct the TFD state corresponds to the configuration with $m_j=1$ for all $j$. Since subsystem $A$ is left unmeasured, a generic measurement record can contain an arbitrary number of entries with $m_j=-1$. Nevertheless, for simplicity, we postselect outcomes with an even number of negative entries, so that all relevant post-measurement states $\ket{\psi_{\bm{m}}}_A$ have positive fermion parity~\footnote{This effectively renormalizes the probabilities by an order-one constant, which we keep implicit.}. Conditioned on each outcome, subsystem $A$ is left in the post-measurement state
  \begin{equation}
  |\tilde{\psi}_{\bm{m}}\rangle_A
  = {}_B\langle \bm{m}|\mathrm{TFD}(t)\rangle,
  \qquad
  |\psi_{\bm{m}}\rangle_A
  = \frac{|\tilde{\psi}_{\bm{m}}\rangle_A}
  {\big\| |\tilde{\psi}_{\bm{m}}\rangle_A \big\|}.
  \end{equation}
  The measurement probability is $p_{\bm{m}}=C_0 \big\| |\tilde{\psi}_{\bm{m}}\rangle_A \big\|^2$, where the order-one constant $C_0$ reflects the postselection. Since $C_0$ is independent of $\bm{m}$, it can be fixed by the normalization of the probability distribution. This set of post-measurement states defines the projected ensemble $\mathcal{E}_{\mathrm{PE}}=\{p_{\bm{m}}p_J,|\psi_{\bm{m}}\rangle_A\}$, where $p_J$ is the distribution of random couplings and the dependence of $|\psi_{\bm{m}}\rangle_A$ on a given disorder realization is kept implicit. 

  \begin{figure}[t]
    \centering
    \includegraphics[width=0.92\linewidth]{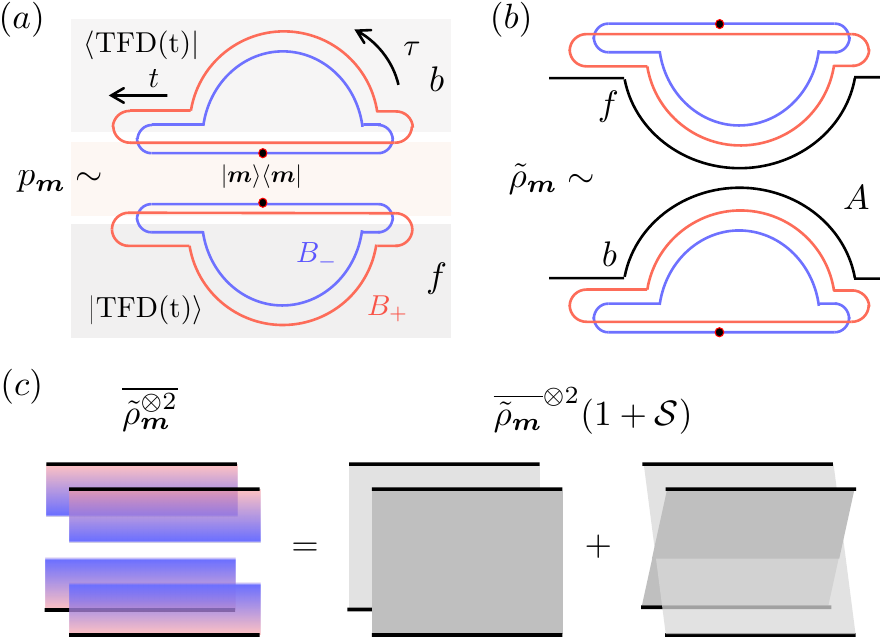}
    \caption{ (a) Path-integral representation of the measurement probability $p_{\bm{m}}$, where the system $B$ is divided into $B_+$ and $B_-$ according to the measurement outcomes. $f/b$ denotes the forward/backward evolution branches. (b) Path-integral representation of the unnormalized post-measurement density matrix $\tilde{\rho}_{\bm{m}}$ for subsystem $A$. (c) Illustration of the two-replica calculation. Here, the colored shaded regions denote the path integral of subsystem $B$, while the gray shaded region indicates the connectivity of the saddle-point solution. }
    \label{fig:path-integral}
  \end{figure}

  Fully characterizing this projected ensemble reveals the emergence of deep thermalization. The central question is whether the ensemble matches the Scrooge ensemble~\cite{PhysRevA.49.668,Goldstein:2005ntv,2008JSP...132..921R} with the same reduced density matrix. Given a density matrix $\sigma$, the Scrooge ensemble is defined as a deformation of the Haar ensemble,
  \begin{equation}\label{eq:Scrooge_def}
  \mathcal{E}_{\mathrm{sc}}(\sigma)=\left\{2^{N_A}\langle\phi|\sigma |\phi\rangle d\phi,  \frac{\sqrt{\sigma}|\phi\rangle}{\sqrt{\langle\phi|\sigma|\phi\rangle}}\right\},
  \end{equation}
  where $d\phi$ is the Haar measure. The Scrooge ensemble achieves the lowest accessible information~\cite{nielsen2010quantum} among all random-state ensembles with the same density matrix and therefore serves as a maximally random ensemble compatible with coarse-grained information about observables. An experimental protocol for its preparation has been proposed using continuous measurements~\cite{Wu:2026svn}.

  \emph{ \color{blue}Measurement probabilities.---} We first analyze the probabilities that weight the projected ensemble using the path-integral formalism. Since $A$ contains only $O(1)$ pairs, its backreaction on the saddle-point equations and on the leading exponential dependence of $p_{\bm{m}}$ is negligible. Thus, at leading order, the measurement probability can be computed by ignoring the unmeasured subsystem $A$, or equivalently by analyzing the path integral for the large measured subsystem $B$. 

  Introducing $Z_{\bm{m}}=Z_\beta \,_A\langle\tilde{\psi}_{\bm{m}} |\tilde{\psi}_{\bm{m}}\rangle_A $, a sketch of the corresponding path-integral contour is shown in Fig.~\ref{fig:path-integral}(a), where the measured subsystem is separated into $B_+$ and $B_-$ according to the signs of the measurement outcomes. The imaginary-time evolution in the TFD state is depicted by a semicircle, while the real-time evolution is shown by straight lines. For a mode $j\in B_+$, taking the overlap with $\ket{m_j=1}_j$ is closely analogous to evaluating the normalization of the TFD state, which imposes the standard antiperiodic boundary condition. In contrast, for modes $j\in B_-$, the antiperiodic boundary condition is replaced by a periodic one. A similar phenomenon has been identified in the evaluation of the Pauli spectrum of thermal density matrices~\cite{Bettaque:2026vpl}. Here, we take an alternative perspective by applying $\ket{-1}_j=\sqrt{2}\chi_j\ket{1}_j$. This allows us to impose antiperiodic boundary conditions also for the subsystem $B_-$, at the cost of additional operator insertions at the black dots in Fig.~\ref{fig:path-integral}(a).

  Next, we evaluate $\overline{p_{\bm{m}}}$ by performing the disorder average. It is established that the SYK partition function is self-averaging~\cite{Kitaev:2017awl}, which guarantees the relation $\overline{p_{\bm{m}}}=C_0\overline{Z_{\bm{m}}}/\overline{Z_\beta}$ in the thermodynamic limit $N\rightarrow \infty$. The partition function $\overline{Z_{\bm{m}}}$ can be analyzed by generalizing standard SYK techniques for equilibrium properties~\cite{maldacena2016remarks}. Leaving the details to the Supplementary Material~\cite{sm}, this leads to the saddle-point equations: 
  \begin{equation}\label{eq:saddle}
  \begin{aligned}
  \Sigma_{ss'}&=J^2f_s f_{s'}\bar{G}_{ss'}^3,\ \ \ G^{B_+}_{ss'}=(\partial_s-\Sigma)^{-1}_{ss'}\\
   G^{B_-}_{ss'}&=G^{B_+}_{ss'}+({G^{B_+}_{Ls}G^{B_+}_{s'0}-G^{B_+}_{Ls'}G^{B_+}_{s0}})/{G^{B_+}_{L0}},
  \end{aligned}
  \end{equation}
  where $G^{B_\pm}_{ss'}$ denotes the two-point function of subsystem $B_\pm$, and $\bar{G}_{ss'}=\lambda G^{B_+}_{ss'}+(1-\lambda) G^{B_-}_{ss'}$ is the two-point function averaged over all modes. Here, we denote the number of fermion pairs in $B_+$ by $N_{B_+}=\lambda N_{B}\approx \lambda N$. The entire contour is parametrized by a single variable $s\in[0,L]$, with $L=\beta+2t$. Under this parametrization, the additional Majorana operators are inserted at $s=0, L$. The auxiliary function $f_s$ takes the values $1$, $i$, and $-i$ when $s$ lies on the imaginary-time, forward real-time, and backward real-time segments, respectively. The saddle-point solution gives
  \begin{equation}
  \begin{aligned}
  -\frac{\ln \overline{Z_{\bm{m}}}}{N}=&\frac{1}{2}\ln \text{det}\left[G^{B_+}\right]+\frac{3}{8}\int dsds'\bar{G}_{ss'}\Sigma_{ss'}\\
  &-(1-\lambda)\ln \Big(2\Big|G^{B_+}_{L0}\Big|\Big),
  \end{aligned}
  \end{equation} 
  which determines $\overline{p_{\bm{m}}}$ given the thermal partition function $\overline{Z_\beta}$. The second line arises from the operator insertions for $B_-$, and the absolute value appears since $B_-$ always contains an even number of fermion pairs. At $\beta=t=0$, this term guarantees $\overline{p_{\bm{m}}}=0$ for any $\lambda<1$. It therefore requires connectivity between the upper and lower branches (see Fig.~\ref{fig:path-integral}(a)) to obtain a finite probability $\overline{p_{\bm{m}}}$. This feature will play an important role in the later analysis of projected ensembles.

  \begin{figure}[t]
    \centering
    \includegraphics[width=0.98\linewidth]{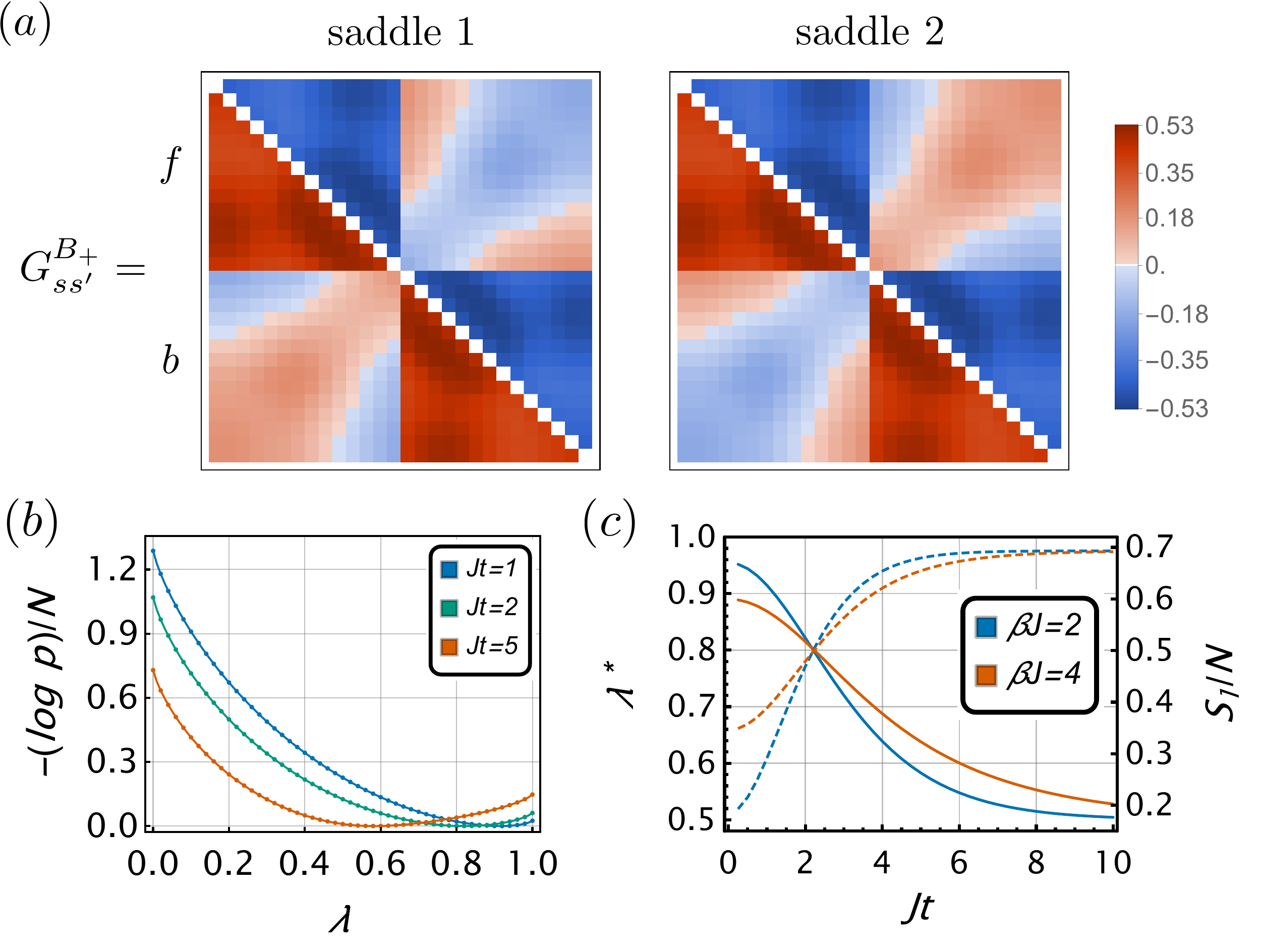}
    \caption{ Numerical results for the measurement probability and participation entropy. (a) Two saddle-point solutions of $G^{B_+}$ for $\beta J=Jt=2$ and $\lambda=0.82$. (b) Probability density $p(\lambda)$ for different values of $Jt$ for $\beta J=2$. (c) The most probable measurement fraction $\lambda^\ast$ and the participation entropy $S_1$ as functions of the evolution time $t$ for $\beta J=2,4$~\cite{note}. Dashed lines correspond to $S_1$, and solid lines correspond to $\lambda^\ast$. }
    \label{fig:num1}
  \end{figure}

  We compute the probability by numerically solving the saddle-point equations, using a method similar to that developed for the subsystem R\'enyi entropy~\cite{Chen:2020wiq,Zhang:2020kia}. An illustration of the corresponding solutions is shown in Fig.~\ref{fig:num1}(a), where the two saddle-point solutions $G^{B_\pm}_{(1/2)}$ are related by flipping the correlation between the upper and lower branches. Consequently, they give the same partition function $\overline{Z_{\bm{m}}}$. Using these solutions, we then compute the measurement probabilities. Since the disorder-averaged probability $\overline{p_{\bm{m}}}$ depends only on the fraction $\lambda$ of positive outcomes, we introduce the continuous distribution $p(\lambda)=N\binom{N}{\lambda N}\,\overline{p_{\bm{m}(\lambda)}}$, where $\bm{m}(\lambda)$ is any record with $\lambda N$ entries equal to $+1$. Thus, $p(\lambda)$ is the disorder-averaged probability density for obtaining a fraction $\lambda$ of positive outcomes and satisfies $\int_0^1 d\lambda\,p(\lambda)=1$. In the limit $N\rightarrow\infty$, we have
  \begin{equation}
  -\frac{\ln p(\lambda)}{N}=-\frac{\ln p_{\bm{m}(\lambda)}}{N}+\lambda \ln \lambda+(1-\lambda)\ln(1-\lambda).
  \end{equation}
  The results of $p(\lambda)$ are presented in Fig.~\ref{fig:num1}(b). For each evolution time $t$, $-\ln p(\lambda)/N$ is a smooth function with a single minimum at $\lambda^\ast$, whose value is zero. The position $\lambda^\ast$ gives the most probable fraction of $+1$ entries in a typical measurement outcome. As $t$ increases, $\lambda^\ast$ shifts from $1$ to $1/2$, with the latter indicating that all measurement outcomes become equally probable. 

  As an immediate diagnostic of these probabilities, we can ask whether the measurement distribution anticoncentrates. We focus on the disorder-averaged Shannon participation entropy $S_1=-\sum_{\bm{m}}\overline{p_{\bm{m}}\ln p_{\bm{m}}}$. In principle, it requires the evaluation of multi-replica probabilities $\overline{p_{\bm{m}}^k}$ and take the limit $k\rightarrow 1$. However, as we will see in the next section, the replica structure of the theory gives $\overline{p_{\bm{m}}^k}=r_k\overline{p_{\bm{m}}}^{k}$ in the limit $N\rightarrow\infty$, where $r_k$ is an order-one constant. Therefore, the participation entropy density $S_1/N$ can then be computed by neglecting $r_n$, rewriting the sum over $\bm{m}$ as an integral over $\lambda$, and applying a saddle-point approximation. For $S_1$, the saddle point is again located at $\lambda^\ast$, yielding 
  \begin{equation}
  {S_1}/{N}=-\lambda^\ast \ln \lambda^\ast-(1-\lambda^\ast)\ln(1-\lambda^\ast).
  \end{equation}
  The result is shown in Fig.~\ref{fig:num1}(c), where $S_1$ smoothly saturates to the maximal value $N\ln 2$. This shows that the dominant measurement records become delocalized over an exponentially large set of outcomes.

  \emph{ \color{blue}Deep thermalization.---}
 With the measurement probabilities understood, we now turn to the projected ensemble by restoring the unmeasured subsystem $A$. We begin with the first moment of the ensemble:
 \begin{equation}
  \rho_A = \sum_{\bm{m}}\overline{p_{\bm{m}}|\psi_{\bm{m}}\rangle_A \langle \psi_{\bm{m}}|}= C_0\sum_{\bm{m}}\rho_{\bm{m},A}.
  \end{equation}
  Here, $\rho_{\bm{m},A}=\overline{|\tilde{\psi}_{\bm{m}}\rangle_A \langle \tilde{\psi}_{\bm{m}}|}$ is the unnormalized density matrix for a given measurement outcome $\bm{m}$. A path-integral representation of $\rho_{\bm{m},A}$ is shown in Fig.~\ref{fig:path-integral}(b). The subsystem $B$ takes the same form as in Fig.~\ref{fig:path-integral}(a), while the uncontracted legs of subsystem $A$ carry the quantum state. Since Eq.~\eqref{eq:saddle} remains valid for the measured subsystem $B$, the density matrix reads~\cite{sm}
  \begin{equation}
  \rho_{\bm{m},A}\sim \frac{\overline{Z_{\bm{m}}}}{\overline{Z_{\beta}}} \sum_a\int \mathcal{D}\chi_A~e^{-\frac{1}{2}\int \sum_{j\in A}\chi_{j}(\partial_s-\Sigma_{(a)})\chi_j}.
  \end{equation}
 Here, we sum the contributions from the two saddle points $a=1,2$. The path integral for subsystem $A$ becomes Gaussian because, in the limit $N_A\ll N$, the interactions experienced by subsystem $A$ arise only through its coupling to the modes in subsystem $B$. 

  Since the self-energy $\Sigma_{(a)}$ depends only on $\lambda$, rather than on the particular outcome $\bm{m}$, we can further denote the path integral on subsystem $A$ as $\hat{\rho}_{A,a}$ and write $\rho_{\bm{m},A}\sim \frac{\overline{Z_{\bm{m}}}}{\overline{Z_\beta}}\sum_a\hat{\rho}_{A,a}(\lambda)$. Furthermore, the two saddle points are related by flipping the correlation between the upper and lower branches. In the operator representation, this corresponds to the action of a parity operator $U=i^{N_A}\prod_{j\in A}\chi_{j}\eta_j$ on subsystem $A$ as $\hat{\rho}_{A,2}(\lambda)=\hat{\rho}_{A,1}(\lambda)U$. Therefore, we find 
  \begin{equation}\label{eq:lambdaint}
  \begin{aligned}
  \rho_{A}&\propto\int d\lambda~p(\lambda)\hat{\rho}_{A,1}(\lambda)(1+U)\propto\hat{\rho}_{A,1}(\lambda^*)P_+,
  \end{aligned}
  \end{equation}
  where $P_+=(1+U)/2$ denotes the projection operator onto the positive-fermion-parity subspace. The appearance of $P_+$ is expected, since we have postselected measurement outcomes with an even number of $-1$ entries~\footnote{For an odd number of $-1$ entries in the measurement outcome, the Majorana insertions in the $B_-$ subsystem give a negative relative sign between the two saddle-point solutions, yielding $\overline{Z}_{\bm{m}}\propto \big(2G_{L0}^{B_+}\big)^{N_{B_-}}$. This results in the factor $(1-U)\propto P_-$.}. The overall normalization is fixed by the condition $\mathrm{tr}[\rho_A]=1$ and is not relevant to the following discussion. In the Supplementary Material~\cite{sm}, we further provide an explicit expression for $\hat{\rho}_{A,1}(\lambda^\ast)$ and verify that it is exactly the reduced density matrix obtained by tracing out subsystem $B$.

  \begin{figure}[t]
    \centering
    \includegraphics[width=0.95\linewidth]{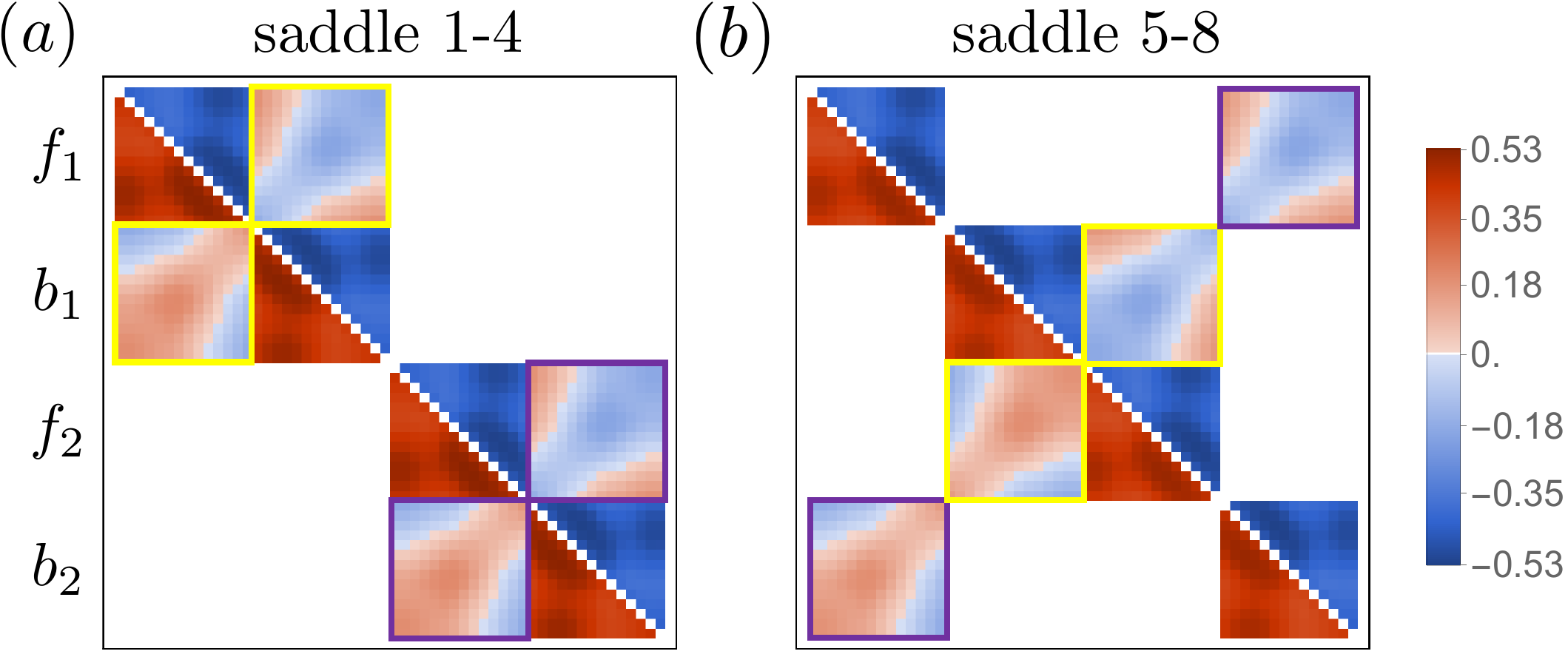}
    \caption{ Saddle-point solutions of $G^{B_+}$ for $\beta J=Jt=2$ and $\lambda=0.82$. We show two representative solutions corresponding to the two distinct pairing patterns illustrated in Fig.~\ref{fig:path-integral}(c). From each representative solution, three additional solutions can be obtained by flipping the signs of the interbranch couplings indicated by the yellow and/or purple squares. We denote the forward/backward evolution branch of the $r$-th replica by ${f/b}_r$. In (a), the pairing pattern is $(f_1,b_1)$ and $(f_2,b_2)$, while in (b), it is $(f_1,b_2)$ and $(f_2,b_1)$.
    }
    \label{fig:num2}
  \end{figure}

  We now proceed to calculate higher moments by generalizing the above analysis. The general moments of the projected ensemble are defined as
  \begin{equation}
  \begin{aligned}
  \rho^{(n)}_A=&\sum_{\bm{m}}\overline{p_{\bm{m}}(|\psi_{\bm{m}}\rangle_A\langle \psi_{\bm{m}}|)^{\otimes n}}=\lim_{k\rightarrow 1}\rho^{(n,k)}_A\\
  \propto&\lim_{k\rightarrow 1}\sum_{\bm{m}}\text{tr}_{n+1}^k\overline{|\tilde{\psi}_{\bm{m}}\rangle_A\langle \tilde{\psi}_{\bm{m}}|^{\otimes k}}.
  \end{aligned}
  \end{equation}
  Here, we have introduced $k$ replicas of subsystem $A$, denoted as $A_1, A_2,\cdots A_k$ and denoted $\text{tr}_{n+1}^k=\text{tr}_{A_{n+1},\cdots A_k}$ for simplicity. The calculation is performed for arbitrary integer $k$, followed by an analytic continuation $k\rightarrow1$. It then requires evaluating a path integral with $k$ forward-evolution branches and $k$ backward-evolution branches. An example with $k=n=2$ is shown in Fig.~\ref{fig:path-integral}(c). The path integral over $B$ can be analyzed in a manner similar to Eq.~\eqref{eq:saddle} and we leave the details to the Supplementary Material~\cite{sm}. Numerical results reveal eight distinct saddle-point solutions, of which two representative examples are illustrated in Fig.~\ref{fig:num2}. These saddles arise from two ingredients: (i) two distinct ways of pairing the forward- and backward-evolution branches, and (ii) four possible signs of the pairing strength. For more general $k$, there are $k!$ pairing patterns and $2^k$ different signs of the pairing strength. Summing up all $2^kk!$ saddle points, we find
  \begin{equation}
  \rho^{(n,k)}_A\propto\sum_{\bm{m}}\frac{\overline{Z_{\bm{m}}}^k}{\overline{Z_\beta}^k}\text{tr}_{n+1}^k\Bigg[\big(\hat{\rho}_{A,1}(\lambda)(1+U)\big)^{\otimes k}\sum_{\mathcal{P}\in S_k}\mathcal{P}\Bigg].
  \end{equation}
  The factor $(1+U)$ arises from the different signs of the pairing strength, and the permutation $\mathcal{P}$ between replicas is summed over all elements of the permutation group $S_k$; see Fig.~\ref{fig:path-integral}(c) for the case $k=n=2$. Taking the limit $k\rightarrow1$ for the partition functions and evaluating the sum over $\bm{m}$ using the saddle-point approximation for $\lambda$ as in Eq.~\eqref{eq:lambdaint}, we obtain
  \begin{equation}\label{eq:moments}
  \rho^{(n)}_A\propto\lim_{k\rightarrow 1}\text{tr}_{n+1}^k\bigg[\rho_A^{\otimes k}\sum_{\mathcal{P}\in S_k}\mathcal{P}\bigg].
  \end{equation}
  The overall normalization can be recovered from the condition $\mathrm{Tr}[\rho_A^{(n)}]=1$. 
  
  Finally, we observe that Eq.~\eqref{eq:moments} exactly matches the moments of the Scrooge ensemble~\cite{PhysRevA.49.668,Goldstein:2005ntv,2008JSP...132..921R}. For $\sigma=\rho_A$, the moments of the Scrooge ensemble can be calculated using a similar replica approach~\cite{mcginley2025scroogeensemblemanybodyquantum}:
  \begin{equation}\label{eq:moments_Scr}
  \rho^{(n)}_{A,\text{Sc}}\propto\lim_{k\rightarrow 1}\text{tr}_{n+1}^k\bigg[\overline{\Big(\sqrt{\rho_A}|\phi\rangle \langle \phi |\sqrt{\rho_A}\Big)^{\otimes k}}\bigg].
  \end{equation}
  Here, the average is taken over Haar-random states. For integer $k$, the standard Weingarten calculus reduces Eq.~\eqref{eq:moments_Scr} to Eq.~\eqref{eq:moments} since $\overline{(|\phi\rangle \langle \phi |)^{\otimes k}}\propto \sum_{\mathcal{P}\in S_k}\mathcal{P}$, establishing the equivalence between the projected ensemble and the Scrooge ensemble at the level of all moments. This demonstrates deep thermalization in the SYK model at arbitrarily short evolution times.

  Our calculation provides a simple physical picture for the emergence of the Scrooge ensemble. In the replicated path integral, the measured subsystem $B$ is much larger than $A$ and therefore determines the dominant saddle. Its saddle-point configurations glue each forward-evolution branch to a backward-evolution branch, but this gluing need not occur within the same replica. Instead, all possible pairings between forward and backward branches are allowed, and these pairings are naturally labeled by permutations $\mathcal{P}\in S_k$. The unmeasured subsystem $A$ then propagates in the background generated by these saddles, so that the gluing pattern of $B$ is inherited as a permutation operator acting on the replicated Hilbert space of $A$. Summing over all saddle points therefore produces precisely the replica-permutation structure characteristic of the Scrooge ensemble.

\emph{\color{blue}Discussion.---}
In this Letter, we have developed a path-integral framework for quantum measurements in solvable SYK models, allowing both measurement probabilities and post-measurement states to be treated analytically in the thermodynamic limit. This approach not only gives access to the full measurement-outcome distribution and participation entropy, but also determines all moments of the projected ensemble. We show that these moments exactly coincide with those of the Scrooge ensemble associated with the averaged post-measurement density matrix, thereby providing an analytic mechanism for the emergence of deep thermalization. This equivalence follows from the saddle-point structure of the replicated measurement path integral, in which the measured subsystem glues forward and backward branches in all possible replica pairings. Overall, our results establish the solvable SYK model as a tractable setting for exploring universality in quantum measurements.

We conclude with a few remarks. First, while our analysis imposes postselection on an even number of negative measurement outcomes, the generalization without postselection is straightforward: one can separately include the contributions from even and odd sectors. The resulting ensemble is a probabilistic mixture of even- and odd-parity Scrooge ensembles, as elaborated in the Supplementary Material~\cite{sm}. Second, it would be particularly interesting to generalize the discussion to other SYK-like models and explore the possibility of a finite-time deep-thermalization transition driven by different saddle-point structures. Finally, it would be important to understand how similar mechanisms emerge beyond solvable models. For example, one may ask whether analogous behavior appears in more realistic systems, such as the Fermi-Hubbard model, perhaps within dynamical mean-field theory.

  \textit{Acknowledgement.}
  We thank Ning Sun for helpful discussions. This project is supported by the Shanghai Rising-Star Program under grant number 24QA2700300, the NSFC under grant 12374477, the Quantum Science and Technology-National Science and Technology Major Project 2024ZD0300101, and the Xuemin Institute of Advanced Studies at Fudan University.

\bibliography{ref.bbl}

\begin{thebibliography}{74}
\bibitem [{not()}]{note}%
  \BibitemOpen
  \href@noop {} {}\bibinfo {note} {In the numerics, we leave a small fraction
  $N_A/N=10^{-2}$ of Majorana modes unmeasured to improve the convergence of
  the numerical solution near $\lambda=1$. We have confirmed that moderate
  changes of this fraction do not lead to any appreciable change in the
  results.}\BibitemShut {Stop}%
\end{thebibliography}

\ifarXiv


\section*{Supplementary material}

# Supplementary Material: Emergence of the Scrooge Ensemble in the Sachdev-Ye-Kitaev Model

Zeyu Liu[1,2] and Pengfei Zhang[2,3,*]

[1] Department of Physics, Zhejiang Normal University, Jinhua 321004, China

[2] State Key Laboratory of Surface Physics & Department of Physics, Fudan University, Shanghai, 200438, China

[3] Hefei National Laboratory, Hefei 230088, China



In this Supplemental Material, we present: (1) a derivation of the saddle-point equations; (2) an explicit verification of the reduced density matrix obtained from the projected ensemble; and (3) a discussion of the projected ensemble without postselection.

## DERIVATION OF SADDLE-POINT EQUATIONS

In this section, we derive the saddle-point equations for the measurement probabilities presented in the main text. We begin with the single-replica case, whose path-integral contour was introduced in the main text. Here, we provide additional details on its parametrization by a single real parameter $s \in [0, L]$, with $L = \beta + 2t$. We start from the branch associated with $|\text{TFD}(t)\rangle$, where $s \in [0, \beta/2]$ denotes the $\beta/2$ imaginary-time evolution, followed by the forward real-time evolution for $s \in [\beta/2, \beta/2 + t]$. We then move to the branch associated with $\langle\text{TFD}(t)|$, where $s \in [\beta/2 + t, \beta/2 + 2t]$ parametrizes the backward real-time evolution. The contour is finally completed by the remaining imaginary-time evolution for $\langle\text{TFD}(t)|$. The standard path-integral representation for Majorana fermions then leads to

$$
\begin{aligned}
\overline{Z_{\boldsymbol{m}}} &= \int dJ_{ijkl} P(J_{ijkl}) \int_{\text{B.C.}} D\chi_j \left[\prod_{l\in B_-} (2\chi_l(L)\chi_l(0))\right] \exp(-S[\chi_j]),\\
S &= \int_0^L ds \left[\sum_j \frac{1}{2}\chi_j \partial_s \chi_j + f_s H[\chi_j]\right].
\end{aligned} \tag{1}
$$

Here, B.C. denotes the standard antiperiodic boundary condition for both subsystems $B_+$ and $B_-$, as discussed in the main text:

$$
\chi_j(L/2^-) = -\chi_j(0^+), \qquad \chi_j(L/2^+) = -\chi_j(L^-). \tag{2}
$$

Next, we perform the integration over Gaussian random variables $J_{ijkl}$, which leads to

$$
\overline{Z_{\boldsymbol{m}}} = \int_{\text{B.C.}} D\chi_j \left[\prod_{l\in B_-} (2\chi_l(L)\chi_l(0))\right] \exp\left(-\int_0^L ds \sum_j \frac{1}{2}\chi_j\partial_s\chi_j + \frac{J^2}{8N^3}\int ds ds' f_s f_{s'} \Big[\sum_j \chi_j(s)\chi_j(s')\Big]^4\right). \tag{3}
$$

To proceed, we introduce bilocal fields following the standard SYK technique [1]. Because of the additional operator insertions in subsystem $B_-$, we define averaged Green's-function fields separately for the subsystems $B_+$ and $B_-$ as

$$
G^{B_+}_{ss'} = \frac{1}{N_{B_+}} \sum_{j\in B_+} \chi_j(s)\chi_j(s'), \qquad G^{B_-}_{ss'} = \frac{1}{N_{B_-}} \sum_{l\in B_-} \chi_l(s)\chi_l(s'). \tag{4}
$$

These relations are imposed by introducing additional self-energy fields $\Sigma^{B_\pm}_{ss'}$ as Lagrange multipliers, leading to the reorganized action

$$
\begin{aligned}
\overline{Z_{\boldsymbol{m}}} &= \int_{\text{B.C.}} D\chi_j DG^{B_\pm} D\Sigma^{B_\pm} \left[\prod_{l\in B_-} (2\chi_l(L)\chi_l(0))\right] \exp\Big(-\tilde{S}[\chi_j]\Big),\\
\tilde{S} &= \int_0^L ds ds' \left[\sum_\pm \sum_{j\in B_\pm} \frac{1}{2}\chi_j(s)\left(\delta(s-s')\partial_s - \Sigma^{B_\pm}_{ss'}\right)\chi_j(s') - \sum_\pm \frac{N_{B_\pm}}{2}\Sigma^{B_\pm}_{ss'} G^{B_\pm}_{ss'} + \frac{J^2 N}{8} f_s f_{s'} \bar{G}^4_{ss'}\right].
\end{aligned} \tag{5}
$$

Here, we have used the averaged two-point function $\bar{G}_{ss'} = \lambda G^{B+}_{ss'} + (1-\lambda) G^{B-}_{ss'}$ for simplicity. We then integrate out the Majorana modes. Importantly, after introducing the bilocal fields, all Majorana modes become decoupled, and their integration can be evaluated by Gaussian integrals. This leads to the $G-\Sigma$ action for the measurement probabilities:

$$\overline{Z_{\boldsymbol{m}}} = \int DG^{B\pm} D\Sigma^{B\pm} \left(2(\partial_s - \Sigma^{B-})^{-1}_{L0}\right)^{N_{B-}} \times \\ \exp\left(\sum_{\pm} \frac{N_{B\pm}}{2} \ln\det\left(\partial_s - \Sigma^{B\pm}\right) - \int_{ss'} \left(\sum_{\pm} \frac{N_{B\pm}}{2} \Sigma^{B\pm}_{ss'} G^{B\pm}_{ss'} - \frac{J^2 N}{8} f_s f_{s'} \bar{G}^4_{ss'}\right)\right). \tag{6}$$

Since the action is proportional to $N$, it can be analyzed using the saddle-point approximation. In particular, the variation with respect to $G^{B\pm}$ receives contributions only from the last two terms, leading to

$$\Sigma^{B+}_{ss'} = \Sigma^{B-}_{ss'} = J^2 f_s f_{s'} \bar{G}^3_{ss'} \equiv \Sigma_{ss'}. \tag{7}$$

This explains why the self-energy field introduced in the main text carries no superscript. Similarly, if we perform variation with respect to $\Sigma^{B+}$, we find $G^{B+}_{ss'} = (\partial_s - \Sigma)^{-1}_{ss'}$. Finally, taking variation with respect to $\Sigma^{B+}$ gives

$$G^{B-}_{ss'} = G^{B+}_{ss'} + (G^{B+}_{Ls} G^{B+}_{s'0} - G^{B+}_{Ls'} G^{B+}_{s0})/G^{B+}_{L0}, \tag{8}$$

where we have imposed the relation $G^{B+} ss' = (\partial_s - \Sigma)^{-1} ss'$. This completes the derivation of the single-replica saddle-point equations presented in the main text. After solving these equations, we obtain the saddle-point prediction for the partition function by substituting the solution into Eq. (6) and neglect the integration over bilocal fields, which leads to Eq. (6) of the main text.

We further briefly explain the calculation of the reduced density matrix, where the subsystem $A$ is kept explicitly. In this case, we add a functional integral over $\chi_{j\in A}$, with free boundary conditions at $s = 0, L/2^+, L/2^-$, and $L$, where quantum states are assigned. Following a similar derivation with additional bilocal field $G^A_{ss'}$ and $\Sigma^A_{ss'}$, the action $\tilde{S}$ becomes:

$$\tilde{S} = \int_0^L ds ds' \left[\sum_{S=B_\pm, A} \sum_{j\in S} \frac{1}{2} \chi_j(s) \left(\delta(s-s')\partial_s - \Sigma^S_{ss'}\right) \chi_j(s') - \sum_{S=B_\pm, A} \frac{N_S}{2} \Sigma^S_{ss'} G^S_{ss'} + \frac{J^2 N}{8} f_s f_{s'} \bar{G}^4_{ss'}\right], \tag{9}$$

and $\bar{G}_{ss'} = \lambda G^{B+}_{ss'} + (1-\lambda) G^{B-}_{ss'} + \frac{N_A}{N} G^A_{ss'}$. Next, we expand the last term to leading order in $N_A/N$. Then, all terms relevant to subsystem $A$ are $O(1)$, while all terms for subsystem $B_\pm$ are $O(N)$. As a result, the saddle-point equations for $B$ remain valid. For subsystem $A$, the action is

$$S_A = \int_0^L ds ds' \left[\sum_{j\in A} \frac{1}{2} \chi_j(s) \left(\delta(s-s')\partial_s - \Sigma^A_{ss'}\right) \chi_j(s') - \frac{N_A}{2} \Sigma^A_{ss'} G^A_{ss'} + \frac{J^2 N}{2} f_s f_{s'} \bar{G}^3_{ss'} G^A_{ss'}\right]. \tag{10}$$

As a result, the integration over $G^A_{ss'}$ can be integrated as a Lagrange multiplier, which imposes $\Sigma^A_{ss'} = \Sigma_{ss'}$. This leads to a Gaussian integral of subsystem $A$ with fixed self-energy $\Sigma_{ss'}$

$$S_A = \int_0^L ds ds' \left(\sum_{j\in A} \frac{1}{2} \chi_j(s) \left(\delta(s-s')\partial_s - \Sigma^A_{ss'}\right) \chi_j(s')\right). \tag{11}$$

Next, we consider the generalization to calculate partition functions with two replicas, defined as $\overline{Z^2_{\boldsymbol{m}}}$. By the same self-averaging argument used in the single-replica case, we have $\overline{p^2_{\boldsymbol{m}}} = C_0^2 \overline{Z^2_{\boldsymbol{m}}} / \overline{Z_\beta}^2$. To derive the saddle-point equations for the replicated path integral $\overline{Z^2_{\boldsymbol{m}}}$, we introduce additional superscript $(\alpha)$ with $\alpha = 1, 2$ that labels different replicas. The partition function then reads

$$\overline{Z^2_{\boldsymbol{m}}} = \int dJ_{ijkl} P(J_{ijkl}) \int_{\text{B.C.}} D\chi^{(\alpha)}_j \left[\prod_{l\in B_-} \left(4\chi^{(1)}_l(L)\chi^{(1)}_l(0)\chi^{(2)}_l(L)\chi^{(2)}_l(0)\right)\right] \exp\left(-S[\chi^{(\alpha)}_j]\right), \\ S = \int_0^L ds \left[\sum_j \frac{1}{2} \chi^{(\alpha)}_j \partial_s \chi^{(\alpha)}_j + f_s H[\chi^{(\alpha)}_j]\right], \tag{12}$$

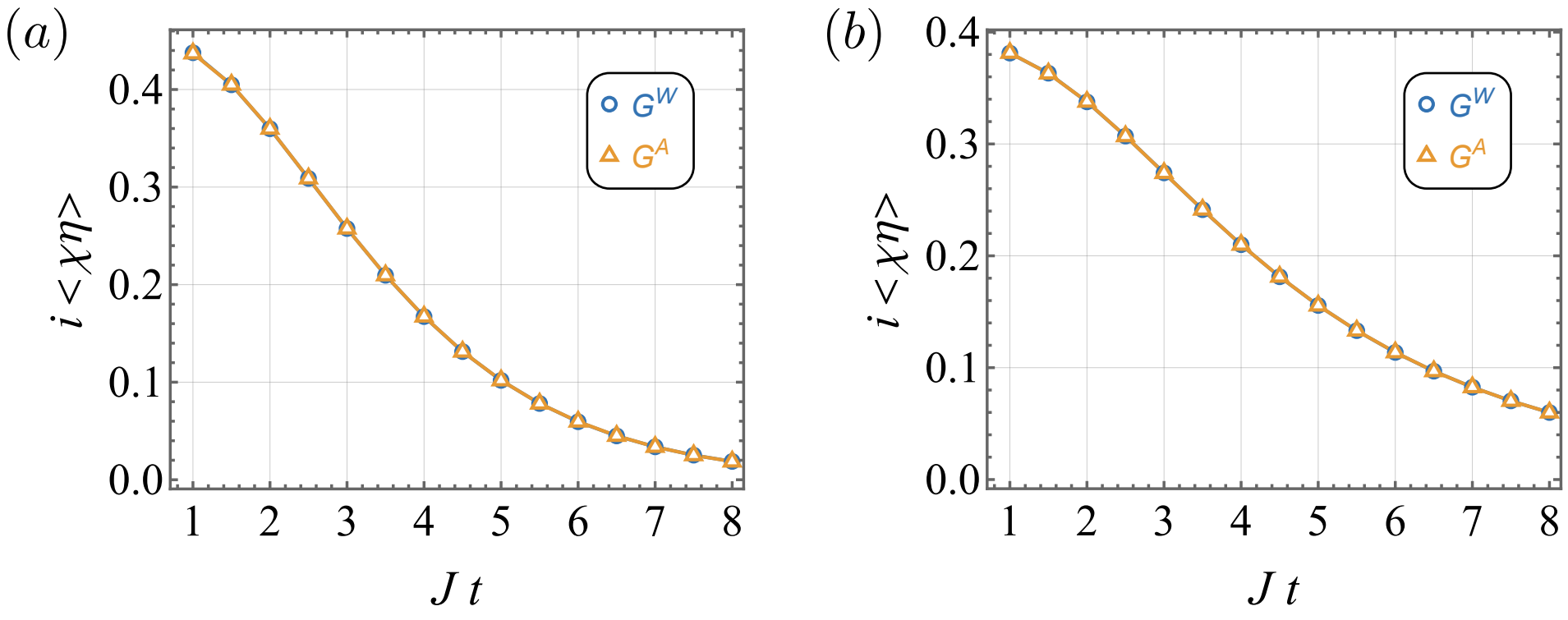


FIG. 1. Numerical verification of the reduced density matrix from the projected ensemble. Here, we choose $\beta J = 2$ and $\beta J = 4$ for panels (a) and (b), respectively.

where the same boundary condition (2) is imposed on all replicas. Next, we perform a similar analysis by integrating out the random couplings and introducing bilocal fields $G^{B\pm,(\alpha\gamma)}_{ss'}$ and $\Sigma^{B\pm,(\alpha\gamma)}_{ss'}$ with additional replica indices $(\alpha,\gamma)$. This leads to

$$\overline{Z^2_{\boldsymbol{m}}} = \int_{\text{B.C.}} D\chi^{(\alpha)}_j DG^{B\pm,(\alpha\gamma)} D\Sigma^{B\pm,(\alpha\gamma)} \left[\prod_{l\in B_-}\left(4\chi^{(1)}_l(L)\chi^{(1)}_l(0)\chi^{(2)}_l(L)\chi^{(2)}_l(0)\right)\right] \exp\left(-\tilde{S}[\chi^{(\alpha)}_j]\right),$$

$$\tilde{S} = \sum_{\alpha\gamma}\int_{ss'}\left[\sum_{\pm}\sum_{j\in B_\pm}\frac{1}{2}\chi^{(\alpha)}_j(s)\left(\delta^{\alpha\gamma}\delta(s-s')\partial_s - \Sigma^{B\pm,(\alpha\gamma)}_{ss'}\right)\chi^{(\gamma)}_j(s') - \sum_{\pm}\frac{N_{B_\pm}}{2}\Sigma^{B\pm,(\alpha\gamma)}_{ss'}G^{B\pm,(\alpha\gamma)}_{ss'} + \frac{J^2N}{8}f_sf_{s'}(\bar{G}^{(\alpha\gamma)}_{ss'})^4\right]. \tag{13}$$

Here, we have introduced the averaged two-point function $\bar{G}^{(\alpha\gamma)}_{ss'} = \lambda G^{B+,(\alpha\gamma)}_{ss'} + (1-\lambda)G^{B-,(\alpha\gamma)}_{ss'}$. The saddle-point equations can be obtained by first integrating out the Majorana fermions and then varying the bilocal fields. Equivalently, one may vary the bilocal fields before integrating out the Majorana modes. We follow the second route. The variation with respect to $G^{B_\pm}$ gives

$$\Sigma^{B_+,(\alpha\gamma)}_{ss'} = \Sigma^{B_-,(\alpha\gamma)}_{ss'} = J^2 f_s f_{s'}(\bar{G}^{(\alpha\gamma)}_{ss'})^3 \equiv \Sigma^{(\alpha\gamma)}_{ss'}. \tag{14}$$

Next, performing the variation with respect to $\Sigma^{B_+}$ gives $G^{B_+,(\alpha\gamma)}_{ss'} = \left(\frac{1}{\partial_s-\Sigma}\right)^{(\alpha\gamma)}_{ss'}$. Finally, the most important difference arises from the variation with respect to $\Sigma^{B_-,(\alpha\gamma)}_{ss'}$. From the expression of $\tilde{S}$, this variation leads to an additional pair of operators $\chi^{(\alpha)}(s)\chi^{(\gamma)}(s')$. As a result, we find

$$G^{B_-,(\alpha\gamma)}_{ss'} = \frac{\langle\chi^{(\alpha)}_l(s)\chi^{(\gamma)}_l(s')\chi^{(1)}_l(L)\chi^{(1)}_l(0)\chi^{(2)}_l(L)\chi^{(2)}_l(0)\rangle}{\langle\chi^{(1)}_l(L)\chi^{(1)}_l(0)\chi^{(2)}_l(L)\chi^{(2)}_l(0)\rangle}. \tag{15}$$

Here, the expectation value is taken in a Gaussian theory for a single mode $l$, defined by the corresponding quadratic action:

$$S_l = \sum_{\alpha\gamma}\int_{ss'}\frac{1}{2}\chi^{(\alpha)}_l(s)\left(\delta^{\alpha\gamma}\delta(s-s')\partial_s - \Sigma^{(\alpha\gamma)}_{ss'}\right)\chi^{(\gamma)}_l(s'). \tag{16}$$

The right-hand side of Eq. (15) can therefore be evaluated using Wick's theorem. These equations are the saddle-point equations for the two-replica system. We solve them numerically to obtain the results shown in Fig. 4 of the main text.

## AN EXPLICIT VERIFICATION OF THE REDUCED DENSITY MATRIX

In this section, we verify our analysis of the projected ensemble by comparing the reduced density matrix obtained from the first moment with the result obtained by directly tracing out subsystem $B$. More explicitly, we hope

to compare $\rho_A \propto \hat{\rho}_{A,1}(\lambda^*)P_+$ as in Eq. (11) of the main text, to $\rho_A \propto P_+\mathrm{tr}_B[|\mathrm{TFD}(t)\rangle\langle\mathrm{TFD}(t)|]P_+$. Here, the projection operator $P_+$ appears because of the post-selection. The equivalence follows directly from $\hat{\rho}_{A,1}(\lambda^*) = \tilde{\rho}_A \equiv \mathrm{tr}_B[|\mathrm{TFD}(t)\rangle\langle\mathrm{TFD}(t)|]$, as we will elaborate below, and from the fact that the projector commutes with arbitrary fermionic density matrix.

As explained in the main text, the density matrix $\hat{\rho}_{A,1}(\lambda^*)$ is a Gaussian state from the perspective of the projected-ensemble calculation. This is also true for $\tilde{\rho}_A$, since in the path-integral representation the only difference arises from the boundary conditions and the insertion of additional operators for $B-$, which do not change the large-$N$ structure. Therefore, checking their equivalence only requires verifying all two-point functions. Since different modes decouples after the saddle-point approximation, an explicit form for the density matrix reads

$$\rho_{A,1}(\lambda^*) = e^{-\kappa\sum_{j\in A} 2i\chi_j\eta_j}/Z, \qquad \tilde{\rho}_A = e^{-\tilde{\kappa}\sum_{j\in A} 2i\chi_j\eta_j}/Z. \tag{17}$$

The only remaining task is to show that $\kappa = \tilde{\kappa}$. This is equivalent to show that the two-point function $\langle i\chi_j\eta_j\rangle$ is then same for both density matrices. For $\tilde{\rho}_A$, this is exactly the Wightman function of the thermal density matrix. Therefore, the relation to check is

$$G^A_{L/2,0} = (\partial_s - \Sigma)^{-1}_{L/2,0} = \mathrm{tr}[i\chi_j\eta_j\rho_{A,1}(\lambda^*)] = \langle\mathrm{TFD}(t)|i\chi_j\eta_j|\mathrm{TFD}(t)\rangle = \langle\chi_j(t - i\beta/2)\chi_j(0)\rangle_\beta = G^W(t). \tag{18}$$

Here, the boundary condition for $A$ on the left-hand side is the standard anti-periodic boundary condition on the entire contour with $s \in [0, L]$. The numerical verification of this relation is presented in Fig. 1, which clearly demonstrate the validity of the relation.

## PROJECTED ENSEMBLE WITHOUT POSTSELECTION

In this section, we discuss the projected ensemble without imposing postselection on measurement records with an even number of negative entries. In this case, the post-measurement state contains quantum states with either even or odd $Z_2$ fermion parity. Nevertheless, since each quantum state has fixed $Z_2$ parity, the projected ensemble cannot be a Scrooge ensemble of the total reduced density matrix $\tilde{\rho}_A = \mathrm{tr}[|\mathrm{TFD}(t)\rangle\langle\mathrm{TFD}(t)|]$. Instead, it is natural to expect the ensemble to be a probabilistic mixture of the Scrooge ensemble with positive fermion parity and that with negative fermion parity:

$$\mathcal{E}_\pm = \left\{ 2^{N_A}\langle\phi|\rho_{A,\pm}|\phi\rangle d\phi, \frac{\sqrt{\rho_{A,\pm}}|\phi\rangle}{\sqrt{\langle\phi|\rho_{A,\pm}|\phi\rangle}} \right\}. \tag{19}$$

Here, we have introduced $\rho_{A,\pm} = \tilde{\rho}_A P\pm/p_\pm$, where $P_\pm = (1 \pm U)/2$ is the projection operator onto the even/odd subspace and $p_\pm = \mathrm{tr}[\tilde{\rho}_A P\pm]$ is the corresponding probability of choosing the ensemble with positive/negative fermion parity. The corresponding moments is given by

$$\rho_A^{(n)} = \sum_\pm p_\pm \mathbb{E}_{\psi\in\mathcal{E}_\pm}(|\psi\rangle\langle\psi|)^{\otimes n}. \tag{20}$$

To show that such an ensemble indeed emerges from the SYK model, we follow the derivation in the main text without requiring $N_{B_-}$ to be even. We still begin with the definition of the moments

$$\begin{aligned} \rho_A^{(n)} &= \sum_{\boldsymbol{m}} \overline{p_{\boldsymbol{m}}(|\psi_{\boldsymbol{m}}\rangle_A\langle\psi_{\boldsymbol{m}}|)^{\otimes n}} = \lim_{k\to1}\rho_A^{(n,k)} \\ &\propto \lim_{k\to1}\sum_{\boldsymbol{m}} \mathrm{tr}^k_{n+1}\overline{|\tilde{\psi}_{\boldsymbol{m}}\rangle_A\langle\tilde{\psi}_{\boldsymbol{m}}|^{\otimes k}}. \end{aligned} \tag{21}$$

We then express the calculation using the path integrals, and derive the saddle-point equations, which takes exactly the same form. The only difference arises when summing up different saddle-points. We need to separate contributions from even and odd $N_{B_+}$, since, as shown in (6), the factor $\left(2(\partial_s - \Sigma^{B_-})^{-1}_{L0}\right)^{N_{B_-}} = (G^{B_+}_{ss'})^{N_{B_-}}$ contributes a minor sign when the inter-branch coupling changes sign if $N_{B_-}$ is even. As a result, we find

$$\rho_A^{(n,k)} = c_0 \sum_{\boldsymbol{m},\,\mathrm{even}N_{B_-}} \frac{\overline{Z_{\boldsymbol{m}}}^k}{\overline{Z_\beta}^k}\mathrm{tr}^k_{n+1}\left[\left(\hat{\rho}_{A,1}(\lambda)(1+U)\right)^{\otimes k}\sum_{\mathcal{P}\in S_k}\mathcal{P}\right] + c_1 \sum_{\boldsymbol{m},\,\mathrm{odd}N_{B_-}} \frac{\overline{Z_{\boldsymbol{m}}}^k}{\overline{Z_\beta}^k}\mathrm{tr}^k_{n+1}\left[\left(\hat{\rho}_{A,1}(\lambda)(1-U)\right)^{\otimes k}\sum_{\mathcal{P}\in S_k}\mathcal{P}\right]. \tag{22}$$

The overall coefficients $(c_0, c_1)$ can always be determined from the normalization condition $\text{tr}[\rho_A^{(n)}] = 1$ and the requirement that the first moment is $\tilde{\rho}_A$. In the continuous limit, the $\overline{Z_{\boldsymbol{m}}/Z_\beta}$ becomes a smooth function of $\lambda = N_{B_-}/N$, independent of whether $N_{B_-}$ is even or odd. As a result, we obtain

$$\rho_A^{(n)} = \tilde{c}_0 \lim_{k\to 1} \text{tr}_{n+1}^k \left[\rho_{A,+}^{\otimes k} \sum_{\mathcal{P}\in S_k} \mathcal{P}\right] + \tilde{c}_1 \lim_{k\to 1} \text{tr}_{n+1}^k \left[\rho_{A,-}^{\otimes k} \sum_{\mathcal{P}\in S_k} \mathcal{P}\right]. \tag{23}$$

It is then straightforward to see that this exactly matches the expectation obtained by generalizing the analysis in the main text via a replica trick to (20).

* PengfeiZhang.physics@gmail.com


\fi

\end{document}